\documentclass{PoS}

\title{Astroparticles in Latin America: current status and outlook}

\ShortTitle{Astroparticles in Latin America: current status and outlook}

\author{\speaker{Ivan Sidelnik}\\%
        Centro At\'omico Bariloche, CNEA, CONICET and Instituto Balseiro, Av. Bustillo km 9.5, San Carlos de Bariloche, Argentina\\
        E-mail: \email{sidelnik@cnea.gov.ar}}

\abstract{
The successful installation and operation of the Pierre Auger Observatory in
Argentina has been a milestone in Astroparticle research in Latin America,
generating new regional research opportunities in the field. In this context,
the LAGO project, begun in 2005 with the aim of studying the high-energy
component of gamma ray bursts (GRBs). This observatory consists of different
arrays of water-Cherenkov detectors installed in high altitude mountains
throughout Latin America. Recently, it has demonstrated the feasibility of
conducting studies on the solar modulation of the galactic cosmic ray flux.
Currently more than 80 scientists and students from Mexico, Guatemala,
Colombia, Venezuela, Ecuador, Peru, Bolivia and Argentina are integrated into
the LAGO Collaboration. The high level of regional integration in the
scientific community reached thanks to this kind of major projects, has led to
the recent formation of the CLES (Consorcio Latinoamericano de Experimentos
Subterr\'{a}neos). This organization promotes the creation and installation of
the ANDES Underground Laboratory to be built inside the projected International
Agua Negra tunnel between Argentina and Chile. The ANDES laboratory with over
1750 meters of rock cover, will be the first laboratory of its kind to be
installed in the Southern Hemisphere.  }

\FullConference{10th Latin American Symposium on Nuclear Physics and Applications\\
		 1-6 December, 2013\\
		 Montevideo, Uruguay}

\begin{document}

\section{Introduction}

The study of astroparticle physics has a long tradition in the Latin American
region. After the discovery of cosmic radiation by Victor Hess in 1912, it was
understood that information about the universe is arriving to the earth not
just in the form of light, but also in particles with mass. Astroparticle physics
studies particle interactions by using astrophysical sources. It also studies the
properties of those objects through the particles that we measure at Earth.

In Latin America the story probably starts in the late forties with the
discovery of the mesons by Lattes, Occhialini and Powell in the Chacaltaya
mountains, in Bolivia \cite{lattes}. This meson happens to be the pion, and
this discovery was made thanks to the altitude that Chacaltaya mountains
provide, to measure particles that weren't attenuated by the atmosphere, and
the improvements that Lattes made in the emulsion of the photographic plates.
This led to a nobel prize awarded to Cecil Powell. Another experience in the
study of cosmic rays was carried out by undergraduate students and researchers
from Argentina, on the Aconcagua mountain \cite{roed}. In the early fifties,
using photographic emulsions they travelled to different sites in the Andes
mountains range and measured the track of particles at different altitudes and
latitudes studying the effect of this on cosmic rays. These first experiences
gave birth to facilities, institutes and open divisions at different
Universities around Latin America that started to focus on the study of cosmic
ray, nuclear and high energy particle physics. 

This paper intends to give a brief review of some of the largest ground based
observatories that are running nowadays measuring the properties of cosmic rays
and gamma-ray bursts, and their application to space weather studies. The
Pierre Auger Observatory, the largest cosmic ray detector in the world, located
in Argentina, is described with its latest results in section \ref{sec:auger}.
Section \ref{sec:lago} and \ref{sec:hawc} describe facilities that were
conceived in principle, as gamma-ray bursts observatories, but they can do
studies of space weather phenomena and cosmic ray also: LAGO (that spans over
all Latin America from Mexico to Antarctica) and HAWC, in Mexico, respectively.
There is a new project that intends to measure solar transient events through
the study of gamma-ray burst and cosmic ray up to the knee energies, called PAS,
section \ref{sec:pas}. The last facility described is ANDES, an underground
laboratory that will be built in a tunnel that goes through the Andes mountains
between Argentina and Chile, section \ref{sec:andes}. The final remarks are
given in section \ref{sec:conc}.

\section{The Pierre Auger Observatory}\label{sec:auger}

The Pierre Auger Observatory is a collaboration of more than 500 members coming
from 18 different countries. Is located in Malarg\"ue, at the south of the
Province of Mendoza, Argentina (1400 m a.s.l.).  Its main goals are studies of
the Ultra High Energy Cosmic Rays (UHECR). The Observatory includes different
instruments working all together with a unique design: it was conceived as a
hybrid detector combining two different techniques of measuring UHECR. The
surface detector array (SD) is composed of 1660 water-Cherenkov detectors (WCD)
placed on a triangular grid of 1500 m covering 3000 km$^2$.  Each station is
filled with 12 tons of water and instrumented with three photomultiplier tubes
(PMTs) which detect the Cherenkov light produced by the charged particles that
go through the water \cite{abraham:04}. On the other hand, the fluorescence
detector (FD) comprises 24 optical telescopes grouped in four sites, observing
the atmosphere above the array. Each telescope consists of an 11 m$^2$
segmented spherical mirror that focuses the fluorescence light into a camera
composed of 440 PMTs, that operates during moonless night \cite{abraham:10}.
The synergy between the SD and FD makes possible the calibration of the energy
of SD events. This has an almost 100\% duty cycle, using the calorimetric
energy measured by the FD ($\sim$10\% duty cycle).

There are upgrades being carried out in the observatory towards low energy
cosmic ray measurements. A denser surface array called ``the infill''
\cite{pao:13}, composed of 71 WCD with spacings of 750 m each, covering an area
of $\sim$30,/m$^2$ it was installed and has been running since 2008. This array
enables the capability of measure cosmic rays with energies from
3$\times$10$^{17}$ eV \cite{abreu:11}, extending the dynamic range of the
observatory to 3 orders of magnitude in energy. This is part of the AMIGA
extension that will also have buried muon counters to measure the chemical
composition of cosmic rays by measuring the muon content of the shower in a
independently. Nowadays there are seven of these detectors installed in the
field \cite{pao:13}. There is also a FD upgrade called ``High Elevation Auger
Telescopes'' (HEAT) that looks over the infill site with three telescopes,
maintaining the hybrid concept of the detector.

Since 2004 the Pierre Auger Observatory has detected high quality UHECR events,
performing key measurements of the energy spectrum, mass composition and
anisotropy of this cosmic rays. In spite of that there are many issues still
not understood.

\begin{figure}
\includegraphics[width=.45\textwidth]{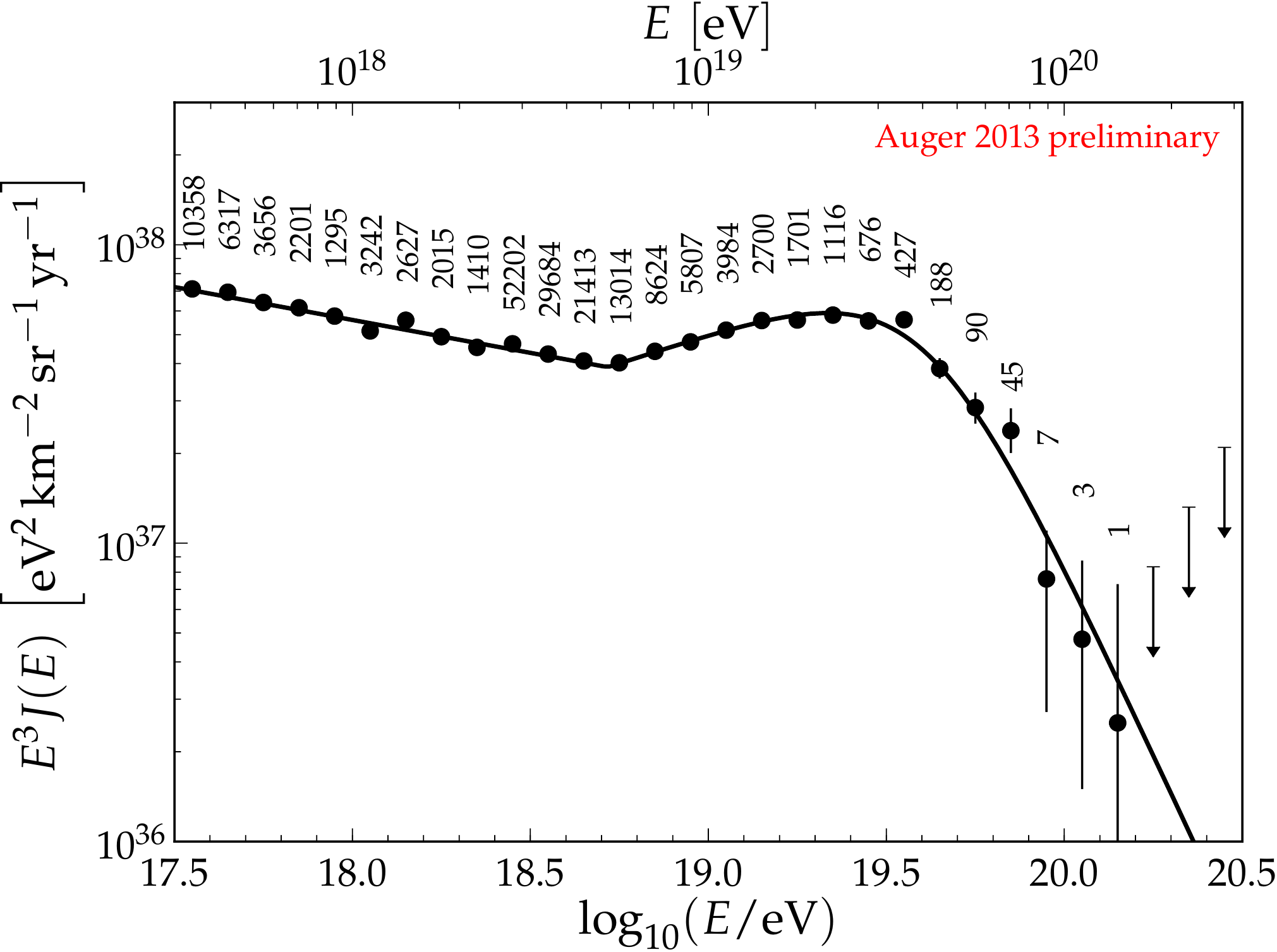}
\includegraphics[width=.55\textwidth]{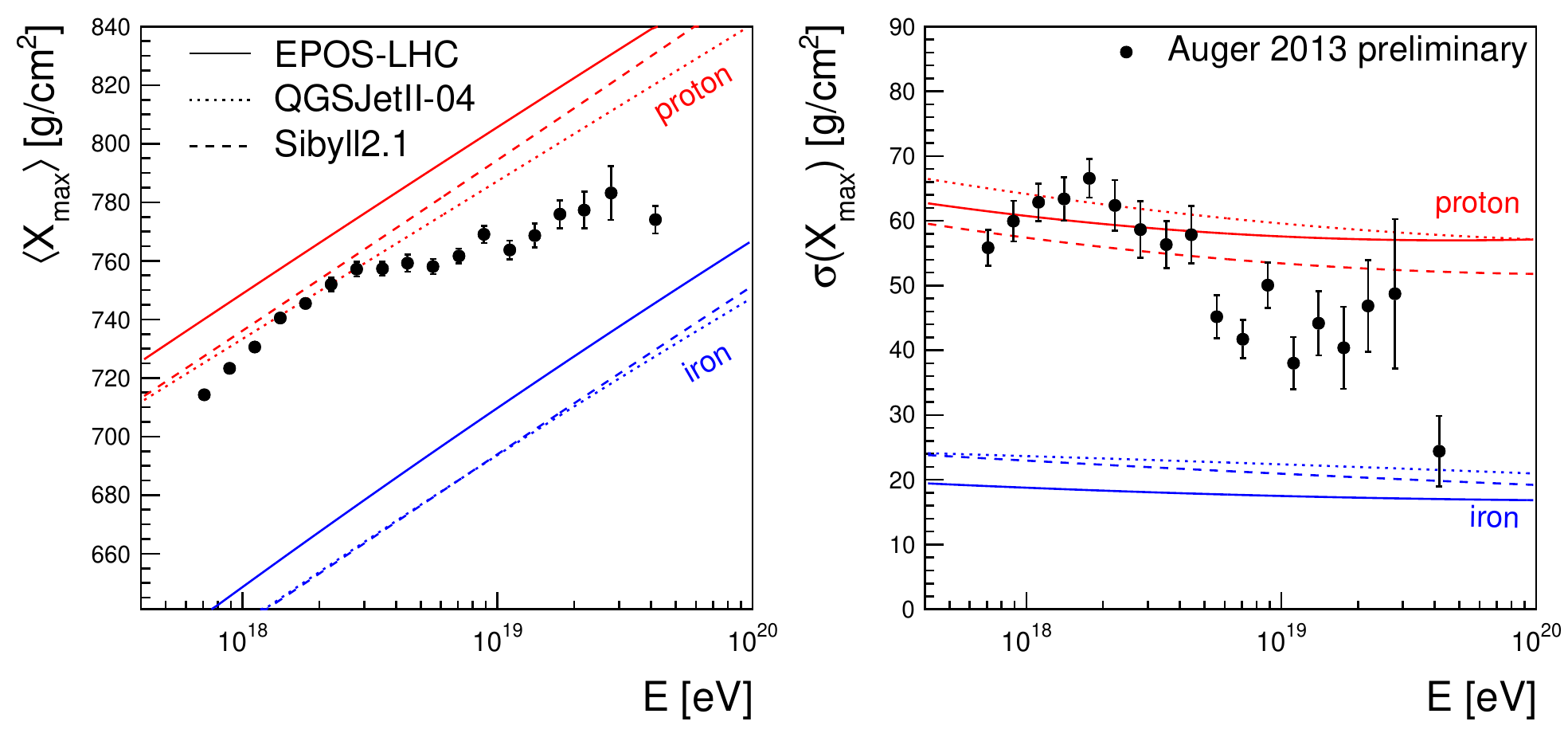}
\caption{Left: Combined energy spectrum of UHECR measured by the Pierre Auger
Observatory. The numbers gives the total events in each energy bin. Right:
change in the mass composition of UHECR as measured by the Auger observatory,
using the mean value and the fluctuation of the X$_{max}$ compared to the
prediction of different hadronic models \cite{pao:13}.} \label{fig:aug1}
\end{figure}

In Fig. \ref{fig:aug1}, left is shown the combined spectrum of the observatory
using SD only, hybrid, inclined and infill measurements. It can be seen to
span over three orders of magnitude in energy \cite{pao:13}. This measurement
shows that the suppression and the ankle are very well established. But despite
its high statistical accuracy, the energy spectrum alone is not enough to
distinguish between the different astrophysical scenarios that can explain
these features. There are too many unknowns like the source distributions and
evolution, the acceleration characteristics, and the mass composition. That is
why this kind of measurements has to be combined with anisotropy and mass
composition measurements.

\begin{figure}
\includegraphics[angle=270,width=.48\textwidth]{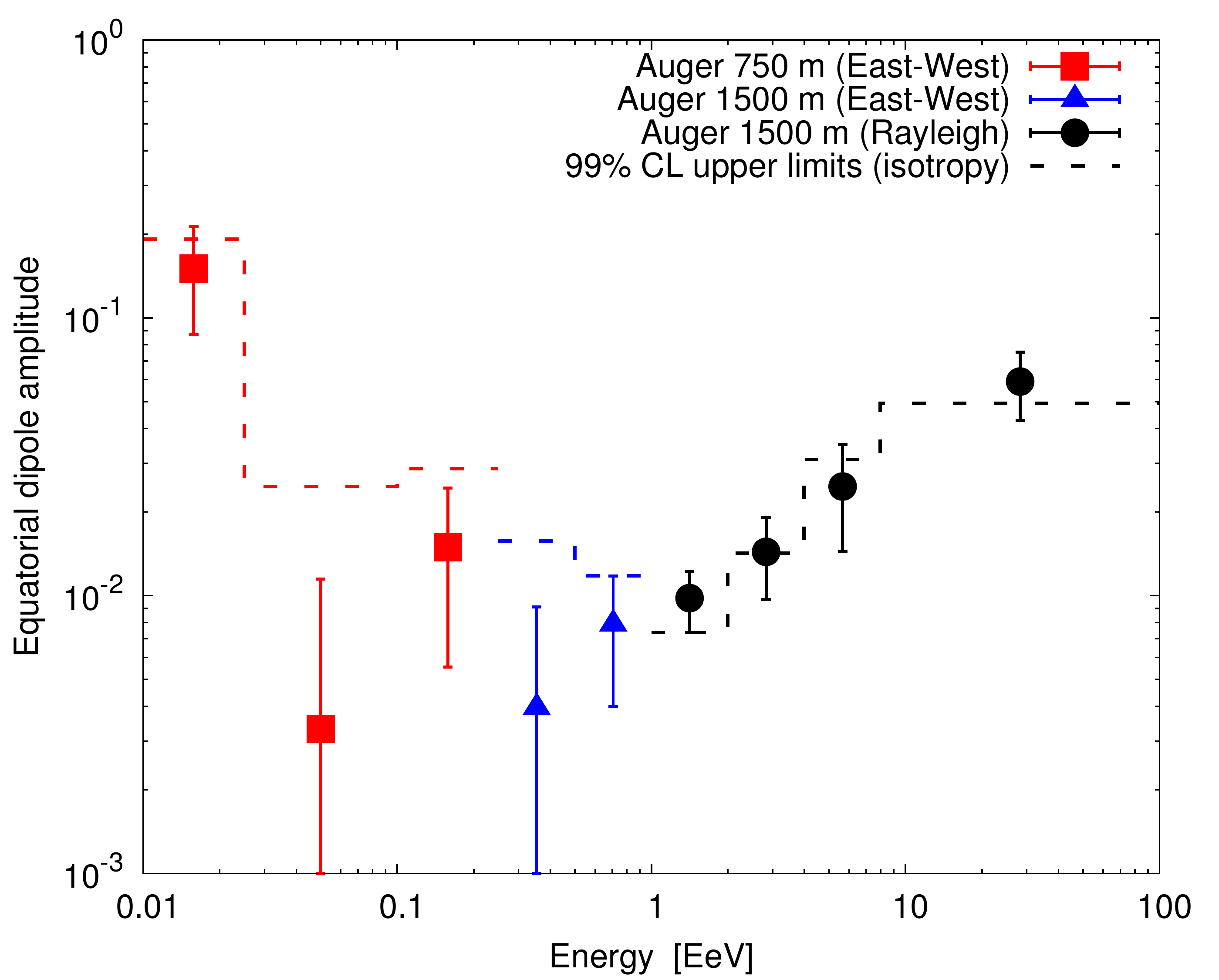}
\includegraphics[angle=270,width=.48\textwidth]{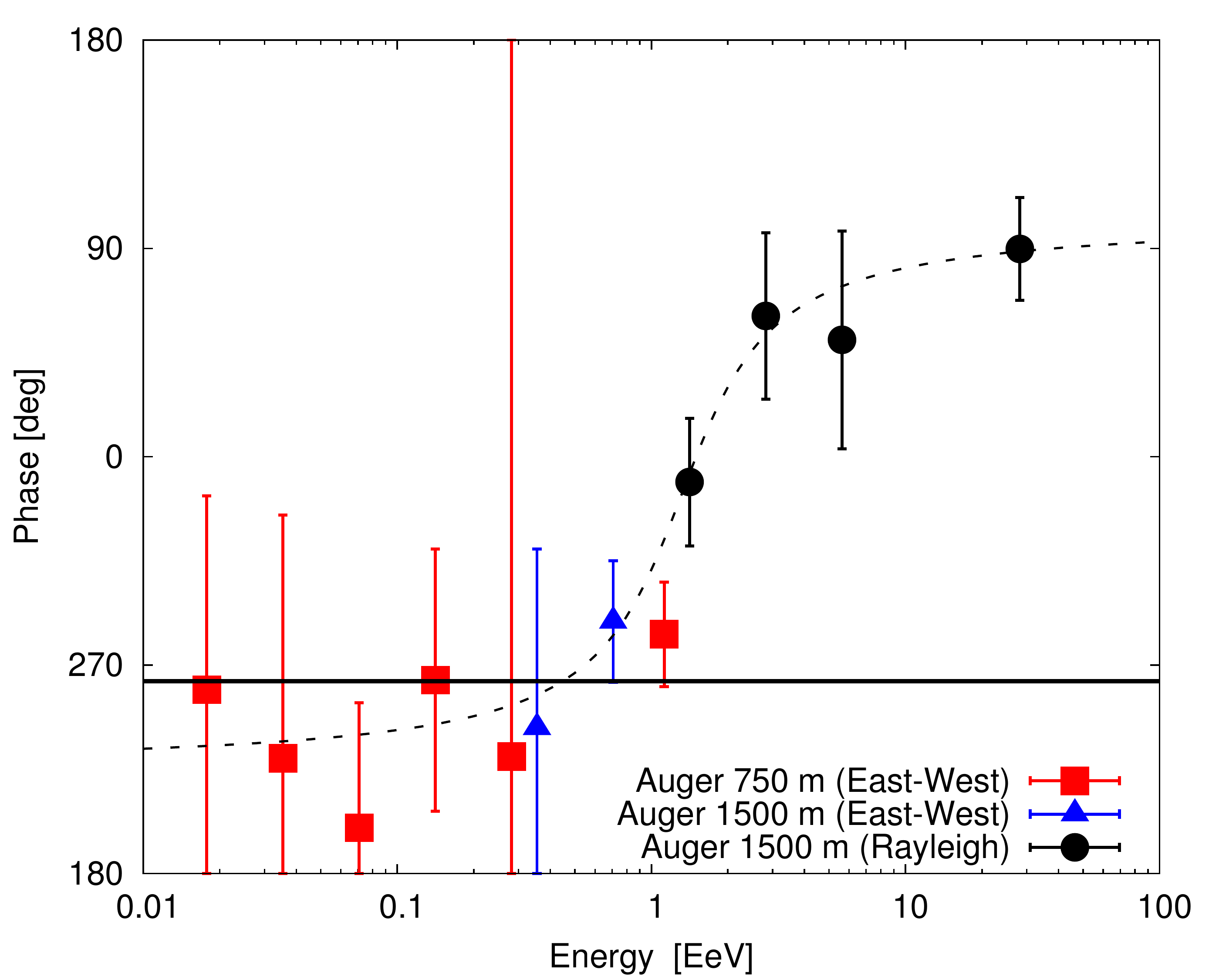}
\caption{Left: Equatorial dipole amplitude as a function of energy.  The dashed lines are the 99\%
CL upper values of the amplitude that could result from fluctuations of an
isotropic distribution. Right: phase of the first harmonic as a function of
energy, it can be seen that there is a phase transition from the galactic
centre towards the anti-galactic centre directions and an alignment at low
energies that could be a hint of large scale anisotropy \cite{pao:13}.
} \label{fig:aug2}
\end{figure}

The results regarding the $<$X$_{max}>$ and $\sigma_{X_{max}}$\footnote{These
are the two first moments of the X$_{max}$ distribution, the position of the
maximum shower size.} and their evolution in energy are shown in Fig.
\ref{fig:aug1}, right. When they are compared to different hadronic models, the
data seems to change the behaviour above the ankle region (1 EeV).

There has also been an analyses on the arrival direction of cosmic rays
performed over the extended energy range, studying the first harmonic
modulation in right ascension. Fig. \ref{fig:aug2}, left, shows the amplitude
of the equatorial dipole that, thanks to the East-West method, spans over four
orders of magnitude in energy. While there is no clear evidence of anisotropy,
it is worth noting how, above 1 EeV, 3 of the 4 points are above the 99\% CL.
On the other hand the phase evolution, Fig. \ref{fig:aug2}, right, shows an
interesting trend with a smooth transition from the galactic centre direction
towards the galactic anti-centre. To test this hypothesis of a change in the
phase there has been set an analyses of an independent data set, such that when
a certain number of events is reached (by the beginning of 2015) this data set
will be ready to confirm or reject the veracity of this result. A summary of
all this contribution and details is available in \cite{ant:13}.

\section{Large Aperture Gamma ray burst Observatory (LAGO)}\label{sec:lago}

Gamma-ray bursts (GRBs) are the most powerful explosions known in the Universe.
They are sudden emissions of gamma-rays lasting very short time intervals (from
0.1 up to 500 seconds). They were discovered in 1967 by the Vela satellites
that the US government launched to monitor covert nuclear tests from space. In
1997 the BeppoSAX satellite detected an optical afterglow of a GRB.  When it
was analyzed it showed that the redshift of the galaxy that originated the GRB
was at more than 6 billon light years from Earth. This showed GRBs to be
extragalactic events. The astrophysical sources of these events are still
unclear but good candidates are the coalescence of compacts objects for the
short bursts (less than 2 s), and hypernovae, supernovae produced by very
massive stars, for the long bursts (more than 2 s) \cite{me:06}.

Despite satellites observations like SWIFT, FERMI and others, revealing
some questions about the origin and location of GRBs, other questions remain
unanswered, especially in the high energy region, i.e. the spectrum. Up to now
no ground-based experiment has detected a GRB.

The LAGO project is a recent collaboration that comes from the association of
latin american astroparticle researchers. It started in 2005 and it was
designed to survey the high-energy component of GRBs. It is a network of
ground-based WCDs, located at mountain altitudes, above 4500 m.a.s.l., where
the flux of the primary gamma-rays is too low for effective detection by small
area satellite detectors. This collaboration was motivated by the experience of
the Pierre Auger Observatory, and the idea is to install uchhis WCD in 9 latin
american countries: Argentina, Bolivia, Brasil, Colombia, Ecuador, Guatemala, Mexico,
Peru and Venezuela \cite{all:08,all:09b}. More than 75 people integrates the
LAGO community, keeping a close collaboration with researchers from IN2P3 of
France and INFN in Italy. 

\begin{figure}
\includegraphics[width=.5\textwidth]{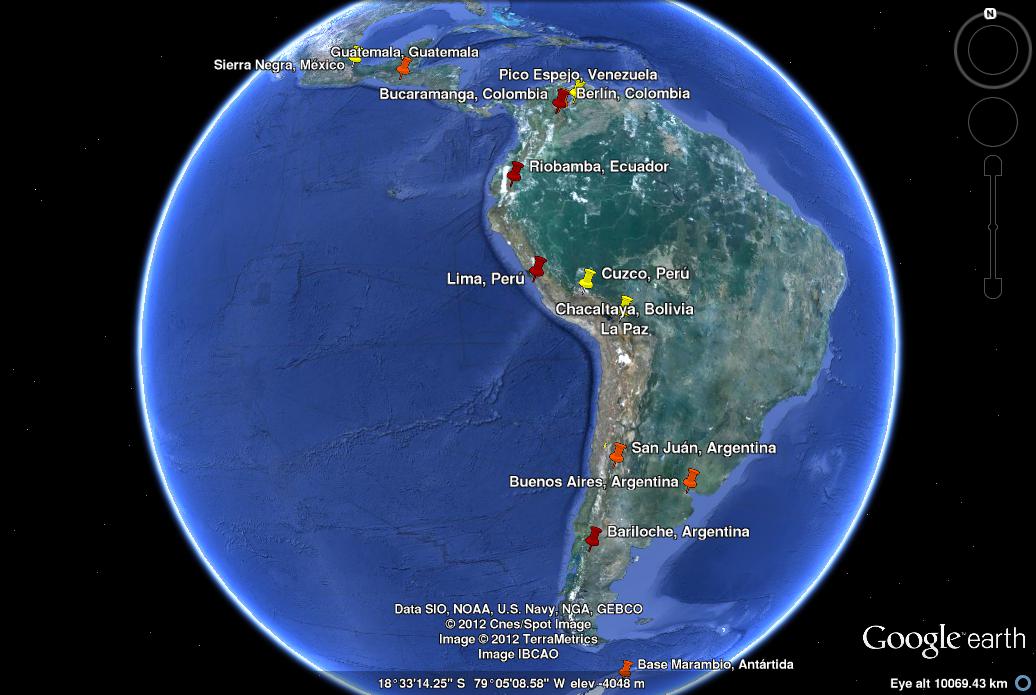}
\includegraphics[width=.5\textwidth]{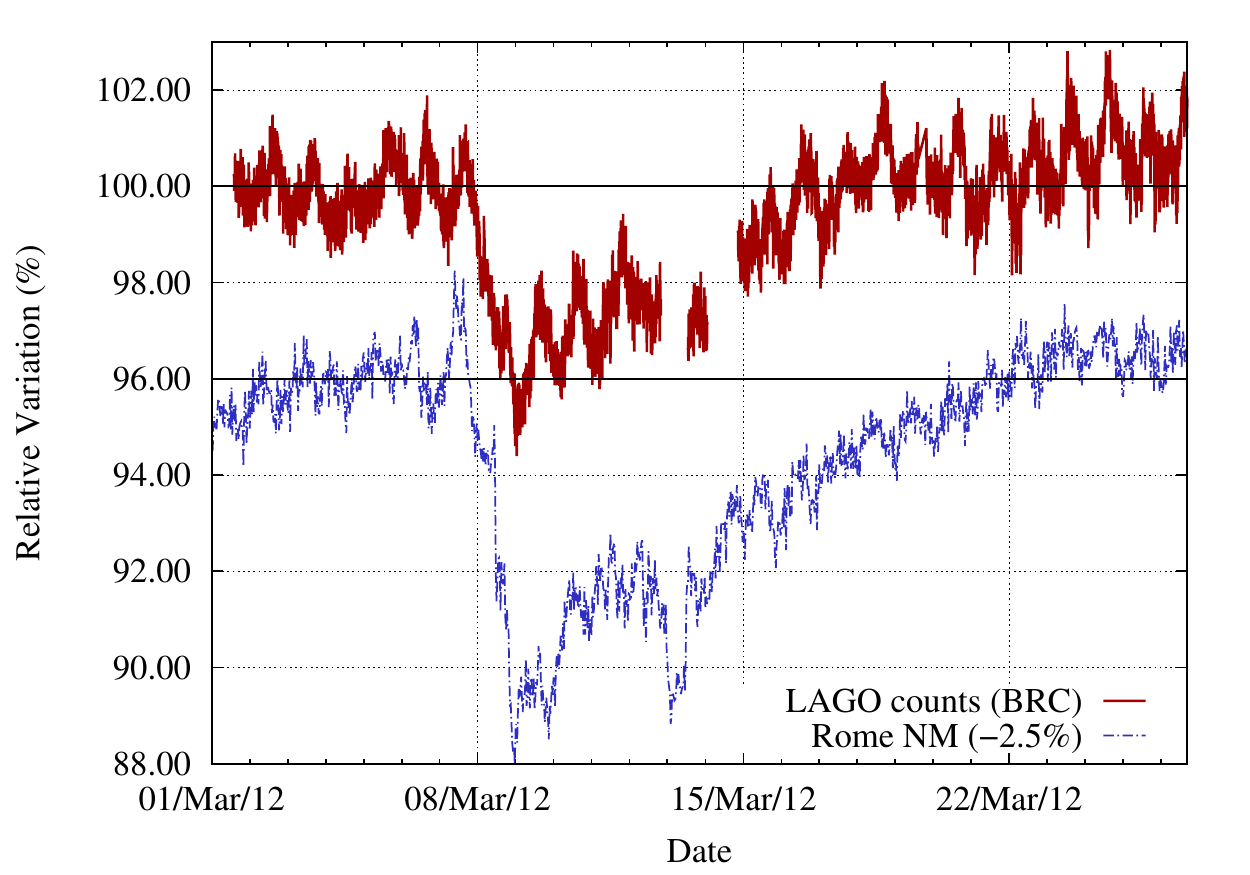}
\caption{Left: Google earth map showing all LAGO sites. For the color code see
text. Right: a Forbush decrease event seen by a single LAGO detector. The
comparison between the WCD in Bariloche and a neutron monitor in Rome shows a very good agreement \cite{asorey:13}.}
\label{fig:lago}
\end{figure}

The idea is that when high energy photons from GRBs reach the atmosphere, they
produce cascades that can be detectable at ground level by using WCDs. Instead
of trying to detect this extensive air showers, LAGO makes use of the single
particle technique, that is to observe an excess in the counting rates of
secondary particles produced in EAS \cite{Vernetto}. The main advantage of
using WCDs compared to other instuments, is their higher sensitivity to
photons, which are 90\% of the secondary particles at ground level for high
energy primary photons.

Although the original plan comprised the detection of the high energy component
of GRBs, recently it has been shown that WCDs can also be used to study the
Solar Modulation (SM) of galactic cosmic rays and other transient effects, by
measuring the variations of the flux of secondary particles at ground level
\cite{pao:11}. 

The main effect is produced by the solar magnetic field. When the Sun shows a
high activity (intense magnetic field), galactic cosmic rays are deflected,
resulting in a reduction of their flux on Earth. When magnetic fields are less
intense we have a higher flux. So, by measuring the flux of galactic cosmic
rays one can determine indirectly the solar activity. This allows, for
instance, the observation of the eleven year cycle of the sun and also
transient events called Forbush decreases. These kind of events are produced
when a Coronal Mass Ejection (CME) is originated in the Sun, resulting in a
huge mass of plasma sent through the interplanetary medium. Upon reaching
Earth, this plasma perturbs the near space and results in a modification in the
flux of galactic cosmic rays. A rapid diminution of their flux can be
observed (a few percent in a few hours). Then, once the CME goes on his way,
the galactic cosmic ray flux slowly comes back to its original value, on a time
scale of days. This reduction in the flux is called Forbush decrease.

LAGO can be an optimal detection network to characterize the SM and transient
events, as it spans over a large area with sites at different latitudes,
longitudes and geomagnetic rigidity cut offs. 

Recently the LAGO Observatory consists of different sites, in
Fig.\ref{fig:lago}, left, a map of the Latin American region can be seen with
the places where a LAGO WCD is in operation or to be installed: Sierra Negra in
Mexico at 4550 m a.s.l. was the first LAGO site and has been in operation since
2007, it has three 2 m$^2$ and two 1 m$^2$ WCD. Chacaltaya in Bolivia at 5250 m
a.s.l. is the highest site of LAGO, and has three 4 m$^2$ and two of 1 m$^2$
WCD. They have been taking data since 2008. Also there is one in Cusco, Peru at
4450 m a.s.l., with one 2 m$^2$ WCD taking data since 2010. There is another
WCD in Bariloche, Argentina, a prototype that was used for calibration and
software tests since 2006, but now has shown that an analyses of SM can be
done. Fig. \ref{fig:lago}, right shows the comparison between a neutron monitor
in Rome, and a single WCD of 1.8 m$^2$ in Bariloche \cite{asorey:13} measuring
a Forbush decrease in 2012. As can be seen the LAGO detector shows similar
relative variation as the other two measurements.

Various other WCDs are installed or being installed: Huancayo and Lima, Peru,
Buenos Aires, Argentina, Cochabamba, Bolivia, and Riobamba, Ecuador are in the
process of installation. Also the cities of Guatemala in Guatemala, Chimborazo
in Ecuador, San Juan with the Marambio base in Artantica (Argentina) are
planned to host WCDs \cite{lago}. There is also a plan to include Brazil
\cite{asorey:14}

\section{The High Altitude Water Cherenkov Observatory (HAWC)}\label{sec:hawc}

HAWC is a survey instrument that aims to study gamma-ray sources in the TeV
region and it is the successor of the Milagro observatory that was located in
New Mexico\cite{mos:13}. It is a joint collaboration between a large number of USA and
Mexican universities and scientific institutions, including UNAM, Los Alamos
National Laboratory and the NASA/Goddard Space Flight Center. The construction
of HAWC started in 2011. It is located at 4100 m a.s.l. in Mexico, near the
Sierra Negra Volcano close to the city of Puebla. It uses WCDs allowing the
observation of cosmic rays also. Each Cherenkov detector consists of 180
m$^3$ of extra pure water inside an corrugated steel tank of 5 m high and 7.3
m of diameter with four PMTs fixed to the bottom of the detector. Currently one
third of the observatory is in operation and taking data.

The HAWC observatory will have an order of magnitude better sensitivity,
angular resolution, and background rejection than the Milagro experiment. The
improved capabilities will allow the detection of both transient and steady
emissions, the study of the Galactic diffuse emission at TeV energies, and it
will measure or constrain the TeV spectra of GeV gamma ray sources. Also, it
will be capable of detecting prompt emission from gamma ray bursts above 50
GeV. A review of this observatory with the status and latest results can
be found in \cite{mos:13}.

\section{Polo de Astronom\'ia Social (PAS)}\label{sec:pas}

Due to its geographic characteristics, the Andinian {\textit{P\'{a}ramo}}
located near Berl\'{\i}n, Colombia, at $3500$\,m a.s.l., is an excellent
location to build an array of particle detectors to study cosmic rays in a wide
energy range, including the solar activity modulation of cosmic rays, gamma-ray
bursts (GRB), and the high energy region of the cosmic ray spectrum. The
proposed array will consist of more than one hundred autonomous and wireless
WCDs located over different concentric triangular grids with different spacing
between neighbor detectors, spanning over a total area of more than
$16$\,km$^2$. This facility will be operated by an interdisciplinary group of
researchers closely related to the LAGO international Collaboration.

The design of the detector array is based on Corsika air shower and Geant4
based detector response simulations. The proposed design of this array will
allow one to implement two different measurement modes: the counting mode and
the shower mode. In the counting mode, the variations in the recorded flux of
secondary particles at detector level can be correlated with transient
phenomena, such as the solar modulation of galactic cosmic rays or the arrival
of the highest energy component of an energetic GRB; or long term flux
modulations related with the solar activity cycle. In the shower mode, in
contrast, there will be a search for time-space correlated signals in
different, non-aligned, detectors of the array. In this way, it will be
possible to determine the main parameters that characterize the extensive air
shower produced by the interaction of a single high energy cosmic ray with the
atmosphere. From this parameters, the arrival direction and the energy of the
impinging cosmic ray will be obtained. The size of the array and the increasing
spacing between detectors will allow one to complement present measurements in
the so called knee region of the cosmic ray energy spectrum ($E \sim
10^{15}$\,eV) and beyond \cite{pas}.

\section{Agua Negra Deep Experiment Site (ANDES)}\label{sec:andes}

In the line of big astroparticle experiments that were built in Latin America
during the pasts years, a new facility is being developed. Deep underground
laboratories are one of the principal contributors to astroparticle physics. In
the past sixty years they provided unique ways of studying weakly interactive
particles. Nowadays more than a dozen underground laboratories are running or
being constructed, studying subjects such as low radioactivity background
measurements, neutrino physics or dark matter searches. All of these laboratories are
located in the northern hemisphere. A review of underground facilities
around the world can be found in \cite{smith:11}.

In the past, some attempts were made to install underground laboratories in the
southern hemisphere, but none of them were maintained. There was one in a gold
mine in South Africa that contributed to the discovery of atmospheric neutrinos
in 1965, another installed in an iron mine in Argentina in 1995 to seek for
dark matter oscillation signals, and also there were unsuccessfully searches
performed for mines in Brazil and Chile, see \cite{be:13} and references in
there. 

A demand for a southern hemisphere laboratory has been growing since different
experiments have claimed that they observed dark matter signals. DAMA/LIBRA has
claimed the observation of a yearly modulation of their signals and attributed
it to dark matter. To understand if this modulation is authentic, an experiment
in the southern hemisphere observing the same kind of modulation is needed. If
an opposed modulation is measured, this could indicate that the signal is
coming from an atmospheric effect.
 
The Andes mountains are a natural limit between Argentina and Chile and have
become of strategic economic importance in the region. The Agua Negra tunnel is
the principal project that could improve the connectivity between Argentina and
Chile. In 2012 the final project for this tunnel was proposed, and it is
planned to be constructed between San Juan and Coquimbo. Also in March 2012 the
presidents of Argentina and Chile gave the green light for the public tender.
This process started in January 2013, and the international call for companies
to compete in the tender was issued in June 2013. The tender process will last
up to end 2014, given the difficulties of the bi-national civil work. The
construction of the tunnel should start in 2015, and last for 7 years.

\begin{figure}
\includegraphics[width=.57\textwidth]{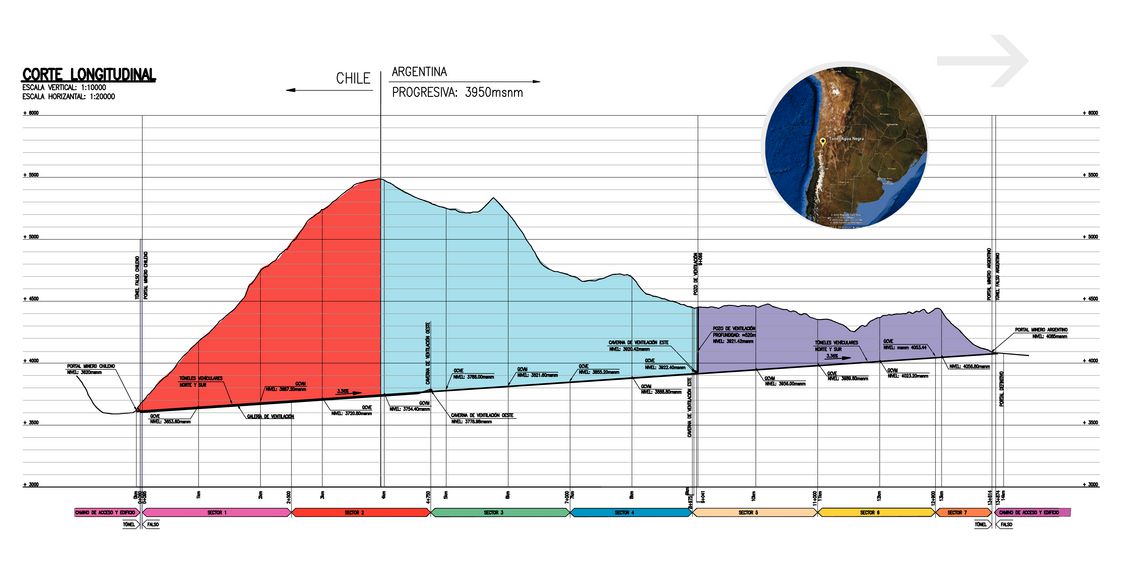}
\includegraphics[width=.45\textwidth]{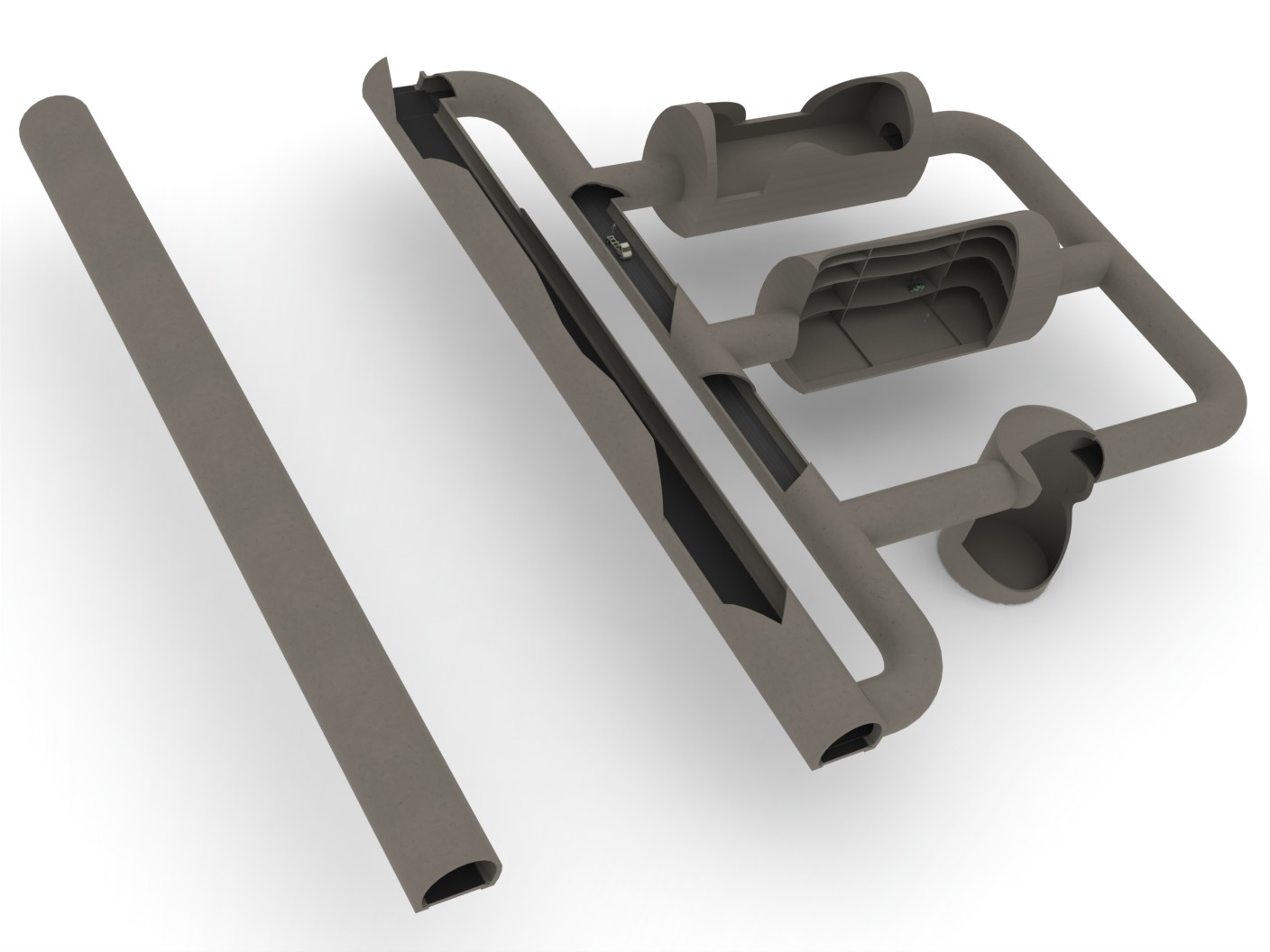}
\caption{Left: Longitudinal cut of the Agua Negra tunnel and its location.
Right: conceptual view of the ANDES deep underground laboratory, it will be
located around 4 km inside the Agua Negra tunnel.} \label{fig:tunel}
\end{figure}

The tunnel design consists of double 14 km long road tunnel, each 12 m of
diameter, separated by 60 m. The entry points are at high altitude, 4100 m
above sea level on the Argentine side, 3600 m on the Chilean side. Most of the
ventilation for the tunnel is natural, thanks to the slope, with a forced
ventilation system used in case of emergencies. The deepest point of the tunnel
is located below the international limit between both countries, at about 4 km
from the Chilean entry. With 1750 m of rock overburden, this spot is ideal to
host a deep underground laboratory. Fig. \ref{fig:tunel}, left, shows a cut in
the Andes mountains were the tunnel is going to be constructed.

Given the relatively high altitude of the laboratory and its international
location, two support laboratories are planned for ANDES. The Argentine
laboratory is expected to be in Rodeo, a small town at 60 km from the tunnel
entrance. It will be the closest support laboratory and be mostly used for day
to day activities and the running of the experiments in ANDES. The Chilean
laboratory will be located in La Serena, at 180 km from ANDES but in an
internationally connected city, with strong scientific presence (such as the
ESO for example). It will be mostly used for the preparation of the
installation of the experiments and their testing.

The ANDES underground laboratory itself is foreseen to have a main hall of 21 m
width, 23 m high and 50 m long, to host large experiments, and a big pit of 30 m
of diameter and 30 m of height for a single large neutrino experiment. A
secondary cavern of 16 m by 14 m by 40 m will host smaller experiments and
services, while three smaller caverns (9 m by 6 m by 15 m) will have dedicated
experiments and a 9 m diameter by 9 m height pit will focus on low radiation
measurements. A conceptual layout of the laboratory can be seen in Fig.
\ref{fig:tunel}, right.

The scientific programme of ANDES includes the main topics in astroparticle
physics as neutrino and dark matter. There will be also a low radiation
measurement laboratory, a geophysics laboratory, space for biology experiments,
and possibly a particle accelerator to do nuclear astrophysics. In neutrino
physics, different experiments will be run. The idea is that ANDES could host
part of a large double beta decay experiment such as SuperNEMO. The flag
experiment of ANDES will be a large neutrino detector similar to KamLAND and
Borexino, but at in a 3 kton scale \cite{neutandes}, focusing on low energy
neutrinos \cite{be:13}. This detector would allow complementary observation of
neutrinos from a nearby supernova, something essential to properly study the
effect of matter on neutrino oscillations. It will also be a geoneutrino
observatory. 

ANDES was considered from the beginning as a unique opportunity not only to
build an underground laboratory for the international community but to build
directly an international laboratory. Given its location on the borderline
between two countries, and the current geopolitical unity displayed by Latin
American countries, ANDES was proposed to be run by a consortium of Latin
American countries, the CLES (initials of Latin American Consortium for
Underground Experiments in Spanish or Portuguese). The CLES is currently formed
by Argentina, Brazil, Chile and Mexico, and is foreseen to be open to more
countries. It will be the organ in charge of the installation and operation of
the ANDES deep underground laboratory and its support laboratories. It will
also organise the academic integration of the scientific activities in the
laboratory with the regional systems. The CLES should be a pole for underground
science in the region \cite{be:13,andes}.

\section{Conclusions}\label{sec:conc}

The lasts 10 years have seen the advent of different kind of ground based
experiments to study astroparticle physics in the Latin American region.  The
Pierre Auger observatory has proved to be a very good role model of a
scientific collaboration, with very important results in the ultra high energy
cosmic ray field that nevertheless has left so far many unanswered questions.
Many efforts are being made to upgrade the observatory to enlarge the
capability of measure cosmic rays at lower energy. Also, this collaboration has
given the experience to spin off experiments like LAGO that will be a large
network spanning all over Latin America to measure GRBs and transient solar
events. The installation of HAWC in Mexico has shown a very good starting and
the PAS project a promising future. 

One of the most important things is the future coming of the ANDES underground
laboratory, a unique facility in the southern hemisphere, that will bring the
opportunity to perform different kind of experiments from dark matter
modulation to supernovae neutrinos and geoneutrinos. The construction is hoped
to start in 2015 and it is foreseen as an international laboratory that can
host experiments from all over the world coordinated by the CLES.  Its large
size and important depth will make it competitive with all existing
laboratories.

\section{Acknowledgments}

I would like to thank the organizers of the 10$^{th}$ LASNPA for allow me to
give this invited talk and my colleagues of the Pierre Auger Collaboration for
all this years I've learned so much. I also thanks Xavier Bertou and Hern\'an
Asorey for their advice, the discussions and the review of this manuscript.
Financial support by the Consejo Nacional de Investigaciones Cient\'ificas y
T\'ecnicas (CONICET) and the ITeDA institute via FUNC is also acknowledged.

\end{document}